\documentclass{article}
\usepackage[spanish,english,activeacute]{babel}
\usepackage[latin1]{inputenc}
\usepackage{amsmath,amssymb,amsfonts,latexsym,cancel}
\usepackage[dvips]{graphicx}
\usepackage[all,knot]{xypic}
\usepackage[authoryear,round,comma]{natbib}
\xyoption{arc} 
\DeclareGraphicsExtensions{.pdf,.png,.jpg}
\newtheorem{theorem}{Theorem}[section]
\newtheorem{proposition}[theorem]{Proposition}

\newcommand{\qed}{\hfill \nobreak \ifvmode \relax \else
      \ifdim\lastskip<1.5em \hskip-\lastskip
      \hskip1.5em plus0em minus0.5em \fi \nobreak
      \vrule height0.75em width0.5em depth0.25em\fi}

\newcommand{\Del}[1]{\Delta_{_{^{#1}}}}

\newcommand{\ND}[1]{ \frac{N_{_{^{#1}}}} {\Delta_{_{^{#1}}}} }

\numberwithin{equation}{section}

\title{\bf{B-splines as a Tool to Solve Constraints in Non-Hydrostatic Forecast Model}}
\author{Álvaro Subías Díaz-Blanco
 \footnote{this work was supported by HIRLAM-B project in cooperation with ALADIN during a month stay visiting the Czech Hydrometeorological Institute}\\
 \scriptsize{Agencia Estatal de METerología, Madrid, Spain}\\
 \scriptsize{\it{asubiasd@aemet.es}}}

\begin{document}
\maketitle
\begin{abstract}
Finite elements has been proven to be an useful tool to discretize the vertical coordinate in
the hydrostatic forecast models allowing to define model variables in full levels
so that no staggering is needed \citep{UH}. In the non-hydrostatic case \citep{Bb}
a constraint in the vertical operators appears (called C1) that does not allow to reduce the set of semi-implicit linear
equations to a single equation in one variable as in the analytic case.
Recently vertical finite elements based in B-splines have been used with an iterative method to relax the C1 constraint \citep{SV}.  
In this paper we want to develop properly some representations of vertical operators in terms of B-splines
in order to keep the C1-constraint. An invertibility relation between integral and
derivative operators between vertical velocity and vertical divergence is also presented.
The final scope of this paper is to provide a theoretical framework of development of finite element vertical operators
to be implemented in the ALADIN-HIRLAM nwp system.\\
\end{abstract}


\section{Introduction}
In the hydostatic model finite elements were introduced by \citet{UH} using linear and cubic $B$-splines 
and Galerkin method to build the integral which is the only vertical operator that appear in the set of semi-implicit linear equations,
however, in the non hydrostatic model \citep{Bb} there is also a derivative operator in the semi-implicit equations.
We focus on two sets of operators of nh-model, on one hand the integral and derivative defined on the $[0,1]$ interval
that relate vertical divergence and vertical velocity and that should be invertible to avoid the production of noise
in the case of advection of vertical velocity. The relationship between vertical divergence and vertical velocity is
\begin{equation}
  d := - g \frac{p}{mR_{d}T} \partial_{\eta} w
\nonumber
\end{equation}

on the other hand we deal with vertical integral operators
$\mathcal{G}^{\ast}$, $\mathcal{S}^{\ast}$, $\mathcal{N}^{\ast}$
(that appear also in the semi-implicit linear equations of the hydrostatic model as $\gamma$, $\tau$, $\nu$,
see \citet{SB,Ri}), they are related by the C1 constraint
\begin{equation}
  \mathcal{G}^{\ast} \mathcal{S}^{\ast} \!- \mathcal{G}^{\ast} \!- \mathcal{S}^{\ast} \!+ \mathcal{N}^{\ast} = 0
\nonumber
\end{equation}

the advantage of satisfying this constraint by the discrete operators is that the set of semi-implicit linear equations
can be reduced to a prognostic equation of a single variable. It was done in finite differences approach \citep{Bb}
through a proper setting of $\alpha^{\ast}_{^{l}}$, $\beta^{\ast}_{^{l}}$, $\delta^{\ast}_{^{l}}$
dimensionless shifts of vertical operators. Sadly, the finite element version cannot be done in that way.
An iterative approach to relax the C1 constraint was proposed by \citet{SV}
setting finite element integral and derivative operators with $B$-splines of any order 
computed with the de Boor's recursive formula and using Galerkin method for the projection operators.\\

The main goal of this paper is to provide a tool to build a grid-point
representation of the two sets of linear operators that are related by an analytical constraints
which must be kept in the discretization step of the construction of the
nh-nwp. The procedure of building the operators is such
that for each operator we associate two sets of basis functions one the image of the other.
At the analytic level the constraints of the operators are satisfied. In particular they are guaranteed for all
representations in terms of basis functions. 
Constraints can be considered as conmutative loops in the application diagrams of associated vector spaces. 
For this purpose we work with $B$-splines (first appeared in \citet{Sc}) 
which constitute a very useful tool in applied math and computer aided design (engineering and graphics).
$B$-splines of order $k$ are piecewise polynomials of order $k-1$ with $k-2$ continuous derivatives,
they are defined in a recursive manner starting from characteristic functions which
gives them suitable properties as being a partition of unity.
In particular we are interested on their well-suited analytical properties under integration and derivation
which constitutes the fundamentals of this paper.\\

To build the grid-point version of the continuous linear operators we also need 
new operators to relate the discretized functions with the basis functions
associated to the linear operators (section \ref{cap:projection}),
such are called projection operators and induce the same loop structure
in the application diagrams at grid-point level.\\

The implementation in the Harmonie forecast model of the formulae presented on this paper has some 
technical difficulties due, among other things, to the Gibbs phenomenon that appears in the jumps
of functions built with the finite element technique, specially with the $\xi, \sigma$ basis
functions developed in the construction of $\mathcal{G}^{\ast}, \mathcal{S}^{\ast}, \mathcal{N}^{\ast}$ operators
that are positive and negative values.\\

The structure of the paper is as follows, in section \ref{cap:bsplines}  
we review first definitions and basic properties of $B$-splines in \ref{cap:definitions}
and integral and derivative formulae in \ref{cap:integrals} that will be used in section \ref{cap:solving_constraints},
in \ref{cap:knots} we make a choice of knots as done in \citet{SV}.
Section \ref{cap:solving_constraints} is the main part of this paper where we build the discretized finite element operators
that keep the analytical constraints, in \ref{cap:gridpoint} we deal with integrals and derivatives and in \ref{cap:c1} 
with the vertical integral operators involved in the C1 contraint.
Finally in section \ref{cap:appendix} we show some useful formulae about the
vertical operators present in the semi-implicit linear set of nh-model (\ref{cap:vertical_operators})
and of the basis functions used (\ref{cap:basis})

\section{Overview of B-splines}
\label{cap:bsplines}

\subsection{Definitions}
\label{cap:definitions}
In this section we will develop their basic properties following closely \citet{dB,dBS}
which constitute a useful introduction to $B$-splines, the starting point of their development is the generatrix function
\begin{equation}
  g_{_{^{k}}}(s;t) := \left\{ 
  \begin{array}{ll}
     \theta(s-t) (s-t)^{k-1} & \ \ \ \ \ {\textrm{\scriptsize{$k\geq 1$}}}  \\
     0                       & \ \ \ \ \ {\textrm{\scriptsize{$k<1$}}} 
  \end{array}
  \right.
\end{equation}

where we define a Heaviside-theta-like function $\theta:=\chi_{_{(0,\infty)}}$ 
as the characteristic function of the interval $(0,\infty)$.
\begin{equation}
   \chi_{_{U}}(x) := \left\{ 
   \begin{array}{ll}
     1 & x \in U \\
     0 & x \notin U
   \end{array}
   \right.
   \ \ \ \ \ \ \ \ \ \ \ \ \ 
   {\textrm{\scriptsize{for any subset $ \ U \subset \mathbb{R}$}}}
\end{equation}

let $\mathtt{t} := \{t_{_{^{i}}} \in \mathbb{R} \}_{_{^{i\in\mathbb{Z}}}}$ a bi-infinite sequence of non-decreasing knots $t_{_{^{i}}} \le t_{_{^{i+1}}}$. 
Associated to this sequence $B$-splines are built as the $k$-th divided difference of $g_{_{^{k}}}(s;t)$ in the first variable
\begin{equation}
  M_{_{^{ik}}}(t) :=g_{_{^{k}}}(t_{_{^{i}}},...,t_{_{^{i+k}}};t)
\label{mg}
\end{equation}

where the $k$-th divided difference of a function is defined recursively
\begin{equation}
  f(t_{_{^{i}}},...,t_{_{^{i+k}}}) := \frac{ f(t_{_{^{i+1}}},...,t_{_{^{i+k}}}) - f(t_{_{^{i}}},...,t_{_{^{i+k-1}}}) }{t_{_{^{i+k}}}-t_{_{^{i}}}}
\end{equation}

and the $0$-th divided difference is the value of the function in the $i$-th knot $f(t_{_{^{i}}})$. 
An useful property of divided differences is given by the Leibniz formula 
\begin{equation}
  ( f \cdot  g ) \!\ (s_{_{^{0}}},...,s_{_{^{k}}}) = \sum_{_{^{r=0}}}^{_{^{k}}} f(s_{_{^{0}}},...,s_{_{^{r}}}) g(s_{_{^{r}}},...,s_{_{^{k}}})
\end{equation}

we define the increments of order $k$
\begin{equation}
  \Del{ik} := t_{_{^{i+k}}}\!\!-t_{_{^{i}}}
\end{equation}

and the normalized $B$-splines
\begin{equation}
  N_{_{^{ik}}} := \Del{ik} \!\ M_{_{^{ik}}}
\label{nm}
\end{equation}

which are equivalent to
\begin{equation}
  N_{_{^{ik}}}(t) =g_{_{^{k}}}(t_{_{^{i+1}}},...,t_{_{^{i+k}}};t) -g_{_{^{k}}}(t_{_{^{i}}},...,t_{_{^{i+k-1}}};t) 
\label{ng}
\end{equation}


%

we define also $S_{_{^{k}}} := \langle \{ N_{_{^{ik}}} \}_{_{^{i \in \mathbb{Z}}}} \rangle$
as the linear subspace of functions spanned by the normalized $B$-splines of order $k$.
The Leibniz formula leads to a recursive expression for $B$-splines
\begin{equation}
 \Del{ik} M_{_{^{ik}}} = (t-t_{_{^{i}}}) M_{_{^{i,k-1}}} + (t_{_{^{i+k}}}\!\!-t)  M_{_{^{i+1,k-1}}} 
\end{equation}

and for normalized $B$-splines (from now on $B$-splines)
\begin{equation}
  \boxed{ N_{_{^{ik}}} = (t-t_{_{^{i}}}) \ND{i,k-1} + (t_{_{^{i+k}}}\!\!-t) \ND{i+1,k-1} }
\label{recursive_b_splines}
\end{equation}

where first order are 
$N_{_{^{i1}}} = \chi_{[t_{^{i}},t_{^{i+1}})}$
which implies that they form a partition of unity, by induction it can be proved that it is also true at all orders
\begin{equation}
  \sum_{_{^{i\in\mathbb{Z}}}} N_{_{^{ik}}}(t) = 1
\label{partition_unity}
\end{equation}

in the case $\Del{ik}\!\!\!=\!\!0$ the quotient $N_{_{^{ik}}}/\Del{ik}$ 
is well defined and takes $0$ value because we take into account $t_{_{^{i}}}=...=t_{_{^{i+k}}}$ in (\ref{mg}, \ref{nm}) formulae.
The support of $N_{_{^{ik}}}$ is $[ t_{_{^{i}}}, t_{_{^{i+k}}} ]$,
given $t\in(t_{_{^{i}}},t_{_{^{i+1}}})$
the set of $k$ elements of $B$-splines that have non-zero value at $t$ is
$ \{ N_{_{^{sk}}} \}_{_{^{s=i-k+1}}}^{_{^{i}}} $

\subsection{Integrals and derivatives}
\label{cap:integrals}
The derivation formula of $B$-splines can be found in \citet{dB, dBS}, the starting point is
the derivative of the generatrix function $\partial_{_{^{t}}} g_{_{^{k}}}(s;t)=(1-k) \!\ g_{_{^{k-1}}}(s;t)$,
with the aid of the (\ref{mg}, \ref{nm}, \ref{ng}) formulae and taking divided differences
we can develop the formula of the derivative of $B$-splines
\begin{equation}
  \boxed{ \partial \!\ N_{_{^{ik}}} = (k-1) \left[ \ND{i,k-1} - \ND{i+1,k-1} \right] }
\label{derivative}
\end{equation}

The integration formula can be also found in \citet{dBLS}.
Let's define the following set of integral operators
\begin{equation}
  \mathcal{I} := \int^{_{^{t}}}_{_{^{t_{\infty}}}} \! dt
  \ \ \ \ \ \ \ \ \ \ \ \ \ \ \ \
  \mathcal{J} := \int_{t_{_{^{-\infty}}}}^{_{^{t}}} \! dt
  \ \ \ \ \ \ \ \ \ \ \ \ \ \ \ \
  \mathcal{N} := \int_{_{^{t_{-\infty}}}}^{_{^{t_{\infty}}}} \! dt
\end{equation}

their expression acting on $B$-splines is (see section {\ref{cap:proofs_integrals}})
\begin{equation}
  \boxed{ \mathcal{I} N_{_{^{ik}}} = -\frac{\Del{ik}}{k} \sum_{_{^{s=-\infty}}}^{_{^{i-1}}} N_{_{^{s,k+1}}} }
  \ \ \ \ \ \ \ \ \ \ \ \ \ \ \ \ \ \ \ \ \
  \boxed{ \mathcal{N} N_{_{^{ik}}} = \frac{\Del{ik}}{k} }
\end{equation}

as an excercise of algebra we can recover the invertibility relations of integral and derivative
$\partial \!\ \mathcal{I} f = f$ and $\mathcal{I} \!\ \partial f = f -f(t_{_{^{\infty}}})$
expressed in spline representation.
The integral and derivative operators relate $B$-splines of consecutive order
\begin{equation}
  \xymatrix @C=1.4cm { S_{_{^{k}}} \ar@<-.3ex>@{<-}[r]_{\partial} \ar@<.3ex>[r]^{\mathcal{I}} & S_{_{^{k+1}}} }
\end{equation}







the derivative and integral are closed operators in $S:= \sum_{^{_{k=1}}}^{_{^{\infty}}} S_{_{^{k}}}$ 
which leads to a matrix representation of them in terms of basis functions.
The fact that $B$-splines are a partition of unity implies that the integral $\int^{_{t}}_{^{_{\ast}}}$ 
has an exact representation in terms of $B$-splines for any starting-point of integration.
In particular $\mathcal{J}=\mathcal{N}+\mathcal{I}$ has the representation
\begin{equation}
  \boxed{ \mathcal{J} N_{_{^{ik}}} = \frac{\Del{ik}}{k} \sum_{_{^{s=i}}}^{_{^{\infty}}} N_{_{^{s,k+1}}} } 
\end{equation}

\subsection{Choice of knots on splines defined on a closed interval}
\label{cap:knots}
Let's consider a closed interval $[ t_{_{^{s}}}, t_{_{^{f}}} ] \subset \mathbb{R}$,
we associate the sequence of knots
\begin{equation}
  \mathtt{t} := \{t_{_{^{i}}} \in \mathbb{R} \}_{_{^{i\in\mathbb{Z}}}}
  \ \ \ \ \ \ \ 
  t_{_{^{i}}} = 
  \left\{
    \begin{array}{lll}
      t_{_{^{s}}} & \ \ {\textrm{\scriptsize{$ i \leq 0   $}}} & {\textrm{\scriptsize{boundary knots}}} \\
      t_{_{^{i}}} & \ \ {\textrm{\scriptsize{$ i = 1,...,I$}}} & {\textrm{\scriptsize{internal knots}}} \\
      t_{_{^{f}}} & \ \ {\textrm{\scriptsize{$ i > I      $}}} & {\textrm{\scriptsize{boundary knots}}}
    \end{array} 
  \right.
\label{knotseq}
\end{equation}


The knot sequence at boundaries chosen here is similar as done in \citet{SV} except for the fact that
now we take a bi-infinite set of knots, i.e., repeating boundary knots as necessary,
the condition on internal knots will be regarded in next section.
For knots of multiplicity $k\!+\!1$ (those for which $t_{_{^{i}}}=t_{_{^{i+k}}}$) we have
$g_{_{^{k}}}(t_{_{^{i}}},...,t_{_{^{i+k}}};t) = 0$ which implies $N_{_{^{ik}}} \equiv 0$ for $i \notin \{-k+1,...,I\}$.
The number of nonzero basis functions is $k+I$ so the space of $B$-splines of order $k$ is
$S_{_{^{k}}} =  \langle \{ N_{_{^{ik}}} \}_{_{^{i=-k+1}}}^{_{^{I}}} \rangle$,
now the partition of unity is written in terms of these functions 
\begin{equation}
  \sum_{_{^{i=-k+1}}}^{_{^{I}}} N_{_{^{ik}}} = \chi_{_{^{[ t_{_{^{s}}}, t_{_{^{f}}} ]}}} 
 \label{partition_unity_knots}
\end{equation}

the $I+2k$ involved knots in the construction of $S_{_{^{k}}}$ are $t_{_{^{-k+1}}},...,t_{_{^{I+k}}}$.
It is important about the derivative operator formula (\ref{derivative}) to keep in mind that for this choice of knots
$N_{_{^{-k,k}}}=0$ and $N_{_{^{I+1,k}}}=0$. The integral operators are
\begin{equation}
   \mathcal{I} N_{_{^{ik}}} = -\frac{\Del{ik}}{k} \sum_{_{^{s=-k}}}^{_{^{i-1}}} N_{_{^{s,k+1}}}
   \ \ \ \ \ \ \ \ \ \ \ \ \ \ \
   \mathcal{J} N_{_{^{ik}}} =  \frac{\Del{ik}}{k} \sum_{_{^{s=i}}}^{_{^{I}}}    N_{_{^{s,k+1}}}
\end{equation}

the recursive relation (\ref{recursive_b_splines}) take a simple form when evaluated at the boundaries
\begin{equation}
\begin{array}{cccccclcl}
N_{^{_{-k+1,k}}} (t_{_{^{s}}})
   & \!\!\!=\!\!\!
   & N_{^{_{-k+2,k\!-\!1}}} (t_{_{^{s}}})
   & \!\!\!=\!\!\!
   & ...
   & \!\!\!=\!\!\!
   & N_{^{_{01}}} (t_{_{^{s}}})
   & \!\!\!=\!\!\!
   & 1 \\
N_{^{_{Ik}}} (t_{_{^{f}}})
   & \!\!\!=\!\!\!
   & N_{^{_{I,k\!-\!1}}} (t_{_{^{f}}})
   & \!\!\!=\!\!\!
   & ...
   & \!\!\!=\!\!\!
   & N_{^{_{I1}}} (t_{_{^{f}}})
   & \!\!\!=\!\!\!
   & 1
\end{array}
\end{equation}

so by the partition of the unity property the values of $B$-splines at boundaries are
\begin{equation}
\begin{array}{lcl}
   N_{_{^{ik}}} (t_{_{^{s}}}) & \!\!\!=\!\!\! & \delta_{_{^{i,-k+1}}} \\
   N_{_{^{ik}}} (t_{_{^{f}}}) & \!\!\!=\!\!\! & \delta_{_{^{iI}}} \\
\end{array}
\label{bs_boundaries}
\end{equation}

\section{Solving constraints using B-Splines}
\label{cap:solving_constraints}

\subsection{Projection operators}
\label{cap:projection}

Let's start defining a finite set of levels $\eta_{(1:L)}:=\{\eta_{_{^{l}}}\}_{_{^{l=1}}}^{_{^{L}}}$
and the space of real-valued functions 
\begin{equation}
  \mathcal{F}_{(1:L)}:=\{ f: \eta_{(1:L)} \rightarrow \mathbb{R} \}
\nonumber
\end{equation}

in this section we will build projection operators between the spaces $S_{_{^{k}}}$ and $\mathcal{F}_{(1:L)}$
\begin{equation}
\xymatrix @C=2.0cm{
     \mathcal{F}_{(1:L)} \ar@<.4ex>[r]^{\mathcal{P}_{_{^{k}}}} &
     S_{_{^{k}}}         \ar@<.4ex>[l]^{\mathcal{Q}_{_{^{k}}}} }
\nonumber
\end{equation}

\subsubsection{$\mathcal{P}$ operator}
a projection matrix $\mathcal{P}$ is obtained evaluating $B$-splines at levels
\begin{equation}
  \mathcal{P}_{_{^{lik}}} := N_{_{^{ik}}}(\eta_{_{^{l}}}) 
  \label{P_original}
\end{equation}

being the partition of unity
\begin{equation}
  \sum_{_{^{i\in\mathbb{Z}}}} \!\ \mathcal{P}_{_{^{lik}}} = 1 
  \ \ \ \ \ \ \ \ \ \
  {\textrm{\scriptsize{$l = 1,...,L$}}} 
\end{equation}

for a given level $\mathcal{P}_{_{^{lik}}}$ is not null for $k$ values of $i\in\mathbb{Z}$.
The boundary condition (\ref{bs_boundaries}) is rewritten
\begin{equation}
\begin{array}{lcl}
  \mathcal{P}_{_{^{1ik}}}
    & \!\!\!\!\!=\!\!\!\!
    & \delta_{_{^{i,-k+1}}} \\
  \mathcal{P}_{_{^{Lik}}}
    & \!\!\!\!\!=\!\!\!\!
    & \delta_{_{^{iI}}}
\end{array}
\label{prj_boundaries}
\end{equation}



\subsubsection{$\mathcal{Q}$ operator}
If the number of levels is $L\!=\!I\!+\!k$ we can take the inverse $\mathcal{Q}_{_{^{k}}}:=\mathcal{P}_{_{^{k}}}^{_{^{-1}}}$ if it exists,
this occurs in the case of non repeated sequence of knots and a set of levels consistent according to \citet{SW}
that states that $N_{_{^{l-k,k}}}(\eta_{_{^{l}}})\neq 0$ for all $l=1,...,L$
which is equivalent to $t_{_{^{l-k}}}<\eta_{_{^{l}}}<t_{_{^{l}}}$. A generalization to repeated knots can be found in \citet{KZ,dB2},
in terms of $\mathcal{P}$ matrix is needed that the diagonal has non-zero elements $\mathcal{P}_{_{^{l,l-k,k}}}>0$.
If $\mathcal{Q}$ is the inverse of $\mathcal{P}$ the boundary condition (\ref{prj_boundaries}) implies that
\begin{equation}
\begin{array}{lcl}
  \mathcal{Q}_{_{^{-k+1,lk}}}
    & \!\!\!\!\!=\!\!\!\!
    & \delta_{_{^{1l}}} \\
  \mathcal{Q}_{_{^{Ilk}}}
    & \!\!\!\!\!=\!\!\!\!
    & \delta_{_{^{Ll}}}
\end{array}
\end{equation}

As pointed out by J. Vivoda (personal communication) the numerical aproximation is better if the set of levels is
located as close as possible to the maxima of $B$-spline functions in order to have a more diagonal dominant matrix $\mathcal{P}$ 
that at the same time have a diagonal dominant $\mathcal{Q}$ so each $B$-spline can be "identified" with each level,
the number of internal knots is less than the number of levels so the statement of internal knots by the condition
that the maxima of their related $B$-splines are close to levels is not a well posed problem.
A numerical treatment can be proposed, this can be done with an iterative method minimizing some objective function.
The general construction makes use of variational analysis, given $f\in\mathcal{F}_{(1:L)}$ we
look for an approximation built on the basis functions
$g_{_{^{l}}} = \sum_{^{_{i\in\mathbb{Z}}}} \mathcal{P}_{_{^{lik}}} c_{_{^{i}}}$
that minimizes some functional $\mathtt{S}:\mathcal{F}_{(1:L)}\rightarrow \mathbb{R}$,
the choice of $c_{_{^{i}}}\!$'s is such that it extremizes the functional $\frac{\partial \!\ \mathtt{S}}{\partial c_{_{^{i}}}}=0$.
The least square functional is
\begin{equation}
  \mathtt{S}[f]=\sum_{_{^{l=1}}}^{_{^{L}}} \left[ f_{_{^{l}}} - \sum_{^{_{i\in\mathbb{Z}}}} \mathcal{P}_{_{^{lik}}} c_{_{^{i}}} \right]^{2}
\end{equation}

we can instead consider a Sobolev inner product in order to smooth out the functions in spline space \citep{O'S}.
The associated euler term of the least square method is
\begin{equation}
  \frac{1}{2} \frac{\delta \mathtt{S}[f] }{\delta c_{i}} = 
       \sum_{^{_{j\in\mathbb{Z}}}} \mathtt{A}_{_{^{ijk}}} c_{_{^{j}}}
     - \sum_{_{^{l=1}}}^{_{^{L}}} f_{_{^{l}}} \mathcal{P}_{_{^{lik}}}
\end{equation}

where $\mathtt{A}_{_{^{k}}} := \sum_{_{^{l=1}}}^{_{^{L}}} \!\ \mathcal{P}_{_{^{lik}}}   \!\ \mathcal{P}_{_{^{ljk}}}$ is an inner product,
i.e., it is positive definite which implies that it is diagonalizable with non negative real eigenvalues.
The coefficients in spline space are 
\begin{equation}
  c_{_{^{i}}} = \sum_{_{^{l=1}}}^{_{^{L}}} \mathcal{Q}_{_{^{ilk}}} \!\ f_{_{^{l}}}
\end{equation}

where we introduce the projection matrix
\begin{equation}
  \mathcal{Q}_{_{^{ilk}}} := \sum_{_{^{j\in\mathbb{Z}}}} [\mathtt{A}_{_{^{k}}}^{_{^{-1}}}]_{_{^{ij}}} \!\ \mathcal{P}_{_{^{ljk}}}
\end{equation}

whose key property is
\begin{equation}
  \mathcal{Q}_{_{^{k}}} \mathcal{P}_{_{^{k}}} = 1
\label{PQ}
\end{equation}


\subsection{Integrals and derivatives}
\label{cap:gridpoint}

%
We induce integral and derivative operators $\hat{\mathcal{J}}$, $\hat{\partial}$
in grid-point space with the aid of projection operators as shown in the following diagram, being $\sim$ an equivalence relation
between continuous or discrete functions that differ by a constant.
An equivalent construction can be done also for $\mathcal{I}$ with its equivalent boundary and invertibility properties.
\begin{equation}
\xymatrix @C=3.0cm @R=1.6cm{
     S_{_{^{k}}}
         \ar@<-.5ex>[d]_{\mathcal{P}_{_{^{k}}}}
         \ar@<.5ex>[r]^{\mathcal{J}} &
     S_{_{^{k+1}}}/\!\!\sim
         \ar@<-.5ex>[d]_{\mathcal{P}_{_{^{k+1}}}}
         \ar@<.5ex>[l]^{\partial} \\
     \mathcal{F}_{(0:L)}
         \ar@<-.5ex>[u]_{\mathcal{Q}_{_{^{k}}}}
         \ar@<.5ex>@{.>}[r]^{\widehat{\mathcal{J}}}  &
     \mathcal{F}_{(0:L+1)}/\!\!\sim
         \ar@<-.5ex>[u]_{\mathcal{Q}_{_{^{k+1}}}}
         \ar@<.5ex>@{.>}[l]^{\widehat{\partial}}}
\nonumber
\end{equation}

this definition is the $B$-spline analog of \citet{UH} where the matrix representation $\mathcal{J}$ and $\partial$
in spline space plays the role of the product of the inverse of mass matrix and stiff matrix $\mathcal{A}^{^{_{-1}}} \mathcal{B}$.
Taking into account the relation (\ref{PQ}) between $\mathcal{P}_{_{^{k}}}$ and $\mathcal{Q}_{_{^{k}}}$ operators
we conclude that the induced operators $\hat{\mathcal{J}}$ and $\hat{\partial}$
inherit the relation of invertibility of $\mathcal{J}$ and $\partial$.
Let's take a set of knots as in (\ref{cap:knots}) section with boundaries $t_{^{_{s}}}\!\!=\!0$ and $t_{^{_{f}}}\!\!=\!1$
in view of the starting point of integration of vertical operators, the internal knots should be taken 
as an intermediate case in order that the maxima of $B$-splines of order $k$ and $B$-splines of order $k+1$ are close to levels.
We choose a configuration where the number of levels is $L\!+\!1\!=\!I\!+\!k$ in order to work with square projection matrices
\begin{equation}
  \mathcal{Q}_{_{^{k}}} = \mathcal{P}_{_{^{k}}}^{_{^{-1}}}
  \ \ \ \ \ \ \ \ \
  \mathcal{Q}_{_{^{k+1}}} = \mathcal{P}_{_{^{k+1}}}^{_{^{-1}}}
\label{pqinverse}
\end{equation}

this ensures that $\mathcal{Q}_{_{^{k}}} 1\!=\!1$, taking into account
the partition of unity property of $B$-splines $\mathcal{P}_{_{^{k}}} 1\!=\!1$. These relations 
are needed to define properly $\mathcal{P}$, $\mathcal{Q}$ operators on the quotient spaces obtained by $\sim$
\begin{equation}
  \sum_{_{^{i=-k+1}}}^{_{^{I}}} \mathcal{P}_{_{^{lik}}}=1
  \ \ \ \ \ \ \Rightarrow \ \ \
  \sum_{_{^{l=0}}}^{_{^{L}}} \mathcal{Q}_{_{^{ilk}}}=1
\label{sum1}
\end{equation}

after some computations the integral and derivative operators induced in grid-point space are
\begin{equation}
\boxed{
\begin{array}{lcccl}
   \widehat{\mathcal{N}}_{_{^{n}}}
      & \!\!\!\!\!=\!\!\!\!\!\!
      & \sum_{_{^{i=-k+1}}}^{_{^{I}}} \frac{\Del{ik}}{k}
      & \!\!\! \mathcal{Q}_{_{^{ink}}} & \\
   \widehat{\mathcal{J}}_{_{^{Mn}}}
      & \!\!\!\!\!=\!\!\!\!\!\!
      & \sum_{_{^{i=-k+1}}}^{_{^{I}}} \frac{\Del{ik}}{k}  
      & \!\!\! \mathcal{Q}_{_{^{ink}}}
      & \!\!\!\!\! [ 1 - \sum_{_{^{\mu<i}}} \mathcal{P}_{_{^{M \mu,k+1}}} ] \\
   \widehat{\partial}_{_{^{mN}}}
      & \!\!\!\!\!=\!\!\!\!\!\!
      & \sum_{_{^{i=-k+1}}}^{_{^{I}}} \frac{k}{\Del{ik}}
      & \!\!\! \mathcal{P}_{_{^{mik}}}
      & \!\!\!\!\! [ \mathcal{Q}_{_{^{i N,k+1}}} \!- \mathcal{Q}_{_{^{i-1,N,k+1}}} ] \\
\end{array}}
\end{equation}

where the indexes run
\begin{equation}
\begin{array}{ccll}
{\textrm{\scriptsize{$M,N$}}} 
& \!\!\!\!{\textrm{\scriptsize{$\in$}}}\!\!\!\!
& {\textrm{\scriptsize{$\{0,...,L\!+\!1$}}}
& \!\!\!\!\!\!{\textrm{\scriptsize{$\}$}}}\\
{\textrm{\scriptsize{$m,n$}}} 
& \!\!\!\!{\textrm{\scriptsize{$\in$}}}\!\!\!\!
& {\textrm{\scriptsize{$\{0,...,L$}}} 
& \!\!\!\!\!\!{\textrm{\scriptsize{$\}$}}}
\end{array}
\ \ \ \ \ \ \ \ \ \ \ \ \ \
\begin{array}{ccl}
{\textrm{\scriptsize{$\mu$}}} &
\!\!\!\!{\textrm{\scriptsize{$\in$}}}\!\!\!\! &
{\textrm{\scriptsize{$\{-k\ \ \ \ ,...,I\}$}}}\\
{\textrm{\scriptsize{$i$}}} &
\!\!\!\!{\textrm{\scriptsize{$\in$}}}\!\!\!\! &
{\textrm{\scriptsize{$\{-k\!+\!1,...,I\}$}}}
\end{array}
\end{equation}

given a set of levels such that $\eta_{_{^{0}}}\!\!=t_{_{^{s}}}$ and $\eta_{_{^{L+1}}}\!\!=t_{_{^{f}}}$
we get the boundary values of projections if we consider (\ref{pqinverse})
\begin{equation}
\begin{array}{lclclcl}
  \mathcal{P}_{_{^{0\mu,k+1}}}
    & \!\!\!\!\!=\!\!\!\!
    & \delta_{_{^{\mu,-k}}}
    & \ \ \ \Rightarrow \ \ \
    & \mathcal{Q}_{_{^{-k,N,k+1}}}
    & \!\!\!\!\!=\!\!\!\!
    & \delta_{_{^{0N}}} \\
  { \mathcal{P}_{_{^{L+1,\mu,k+1}}} } 
    & \!\!\!\!\!=\!\!\!\!
    & \delta_{_{^{\mu I}}}
    & \ \ \ \Rightarrow \ \ \
    & \mathcal{Q}_{_{^{IN,k+1}}}
    & \!\!\!\!\!=\!\!\!\!
    & \delta_{_{^{L+1,N}}} \\
  { \mathcal{P}_{_{^{0ik}}} }
    & \!\!\!\!\!=\!\!\!\!
    & \delta_{_{^{i,-k+1}}}
    & \ \ \ \Rightarrow \ \ \
    & \mathcal{Q}_{_{^{-k+1,nk}}}
    & \!\!\!\!\!=\!\!\!\!
    & \delta_{_{^{0n}}} \\
\end{array}
\end{equation}

which after some calculation we check that boundary values of the integrals are as desired
\begin{equation}
  \widehat{\mathcal{J}}_{_{^{0n}}}\!=0
  \ \ \ \ \ \ \ \ \ \ \ \ \ \ \ \ \
  \widehat{\mathcal{J}}_{_{^{L+1,n}}}\!\!=\widehat{\mathcal{N}}_{_{^{n}}}
\label{intderboundaries}
\end{equation}

as a consequence of the partition of unity property we see that constants are well suited 
to the integral operator $\widehat{\mathcal{N}}1=1$ and the derivative operator $\widehat{\partial}1=0$.
By construction the invertibility relations are inherited in grid-point operators,
let $f\in\mathcal{F}_{(0:L)}$ and $g\in\mathcal{F}_{(0:L+1)}$ be arbitrary functions
\begin{equation}
 \hat{\partial} \hat{\mathcal{J}} f = f 
 \ \ \ \ \ \ \ \ \ \ \ \ \ \ \ \ \ \ \
 \hat{\mathcal{J}} \hat{\partial} \!\ g = g - g(0)
\end{equation}

\subsection{Vertical operators in the semi-implicit equations of the nh-model}
\label{cap:vertical_operators1}
The continuous linear system in the semi-implicit equations of the non-hydrostatic forecast model \citep{Bb}
are written in terms of vertically continuous linear operators which we present in this section.
In order to simplify their expression let's redefine the vertical coordinate as $t\in[0,1]$
\begin{equation}
  t := \frac{\pi^{\ast}(\eta)}{\pi^{\ast}_{s}}
\end{equation}

being the jacobian $dt = \frac{m^{\ast}}{\pi_{s}^{\ast}} d\eta$ (where $m^{\ast} := \partial_{\eta}{\pi^{\ast}}$)
so that in $t$ coordinates the vertical derivative operators are
\begin{equation}
\begin{array}{lcl}
  \partial^{\ast}    & \!\!\!:=\!\!\! & t \!\ \partial_{_{^{t}}} \\[0.2cm]
  \mathcal{L}^{\ast} & \!\!\!:=\!\!\! & \partial^{\ast} ( \partial^{\ast} \!+ 1 )
\end{array}
\end{equation}

and the integral operators
\begin{equation}
\begin{array}{lcr}
  \mathcal{G}^{\ast} f & \!\!\!:=\!\!\! & \int^{_{^{1}}}_{_{^{t}}} f \frac{dt}{t}     \\[0.2cm]
  \mathcal{S}^{\ast} f & \!\!\!:=\!\!\! & \frac{1}{t}   \int^{_{^{t}}}_{_{^{0}}} f dt \\[0.2cm]
  \mathcal{N}^{\ast} f & \!\!\!:=\!\!\! & \int^{_{^{1}}}_{_{^{0}}} f dt
\end{array}
\end{equation}

these operators are related by the C1 constraint which is needed in order
to obtain a semi-implicit linear system depending on a single variable,
this constraint can be proven by a per part integration 
\begin{equation}
  \mathcal{G}^{\ast} \mathcal{S}^{\ast} \!- \mathcal{G}^{\ast} \!- \mathcal{S}^{\ast} \!+ \mathcal{N}^{\ast} = 0 
\end{equation}

\subsection{Resolution of C1 constraint}
\label{cap:c1}
The only step that doesn't guarantee that the space of $B$-splines of all order $S$ is closed
by the action of the vertical operators is the division by $t$,
by virtue of (\ref{recursive_b_splines}) the quotient $N_{_{^{ik}}}/t$ can be split into a $S$-term and a rest 
which is linear combination of $\chi_{[t_{^{_{i}}},t_{^{_{i+1}}})}/t$. 
In the following we will focus on linear spaces of basis functions in which
the vertical operators have closed representation,
for our purposes we factorize the C1 constraint
\begin{equation}
  \boxed{(\mathcal{G}^{\ast} - 1) (\mathcal{S}^{\ast} -1) = (1 - \mathcal{N}^{\ast})}
\label{factorizationC1}
\end{equation}

rewritting $\mathcal{S}^{\ast}=\frac{1}{t} \mathcal{J}$
we seek new basis functions $\xi$ such that $\mathcal{J} \xi_{_{^{ik}}} := t N_{_{^{ik}}}$
in order to ensure $\mathcal{S}^{\ast} \xi_{_{^{ik}}} = N_{_{^{ik}}}$
to relate two basis through $\mathcal{S}^{\ast}$ operator
\begin{equation}
 \xymatrix @C=1.2cm @R=0cm{
            H_{_{^{k}}}     \ar[r]^{\mathcal{J}}  &
            tS_{_{^{k}}}    \ar[r]^{t^{-1}}          & 
            S_{_{^{k}}}      \\
            {\textrm{\scriptsize{$\xi_{_{^{ik}}}$}}}  \ar@{|->}[r] &  
            {\textrm{\scriptsize{$t N_{_{^{ik}}}$}}}  \ar@{|->}[r] &
            {\textrm{\scriptsize{$N_{_{^{ik}}}$}}}     } 
\end{equation}

where we define the vector space $H_{_{^{k}}} := \langle  \{ \xi_{_{^{ik}}} \}_{_{^{i\in\mathbb{Z}}}} \rangle$.
Taking into account that $\partial^{\ast}\!+1$ is the inverse of $\mathcal{S}^{\ast}$
(see appendix \ref{SDinverse}) we write
\begin{equation}
  \xi_{_{^{ik}}} = (\partial^{\ast}\!+1) \!\ N_{_{^{ik}}} = \partial \!\ (tN_{_{^{ik}}})
  \label{xirel}
\end{equation}

the product $tN_{_{^{ik}}}$ belongs to spline space, it can be proved using the recursive equation 
of $B$-splines (\ref{recursive_b_splines}) and reindexing one of the infinity sums
\begin{equation}
  \sum_{_{^{s=i}}}^{_{^{\infty}}} N_{_{^{s,k+1}}} = (t-t_{_{^{i}}}) \ND{ik} + \sum_{_{^{s=i+1}}}^{_{^{\infty}}} N_{_{^{s,k}}}
\end{equation}

once it is derived it is enough to isolate the $tN$-term
\begin{equation}
  tN_{_{^{ik}}} = t_{_{^{i}}} N_{_{^{ik}}} + \Del{ik} \sum_{_{^{s=i}}}^{_{^{\infty}}} \left[ N_{_{^{s,k+1}}} \!- N_{_{^{s+1,k}}} \right]
\label{tn}
\end{equation}

taking derivatives we see that the $\xi$ basis has the expression
\begin{equation}
   \xi_{_{^{ik}}} = k N_{_{^{ik}}} + (k-1) \left[ t_{_{^{i}}} \ND{i,k-1} - t_{_{^{i+k}}} \ND{i+1,k-1} \right]
\end{equation}

a consequence of the fact that the multiplication by a polynomial is a closed operation in $S$
is that $\partial^{\ast}$ and $\mathcal{L}^{\ast}$ are also closed on $S$.
As a consequence of the fact that $B$-splines are partition of unity (\ref{partition_unity}) 
and applying ({\ref{xirel}}) we check that constants can be written as a linear combination of $\xi$ basis
\begin{equation}
  \sum_{_{^{i\in\mathbb{Z}}}} \xi_{_{^{ik}}} = 1
\label{normalizacion_chi}
\end{equation}

in the same way as was done for $\hat{\mathcal{J}}$ and $ \hat{\partial}$
we take a set of knots as in (\ref{cap:knots}) with $t_{^{_{s}}}\!\!=\!0$ and $t_{^{_{f}}}\!\!=\!1$.
The operator relation $ \mathcal{N}^{\ast} (\partial^{\ast}\!+1) = \Pi_{_{^{1}}}^{\ast}$ acting to $B$-splines
(see appendix \ref{cap:vertical_operators2}) and
the boundary condition $\Pi_{_{^{1}}}^{\ast} N_{_{^{ik}}} = \delta_{_{^{iI}}}$ leads to
\begin{equation}
  \mathcal{N}^{\ast} \xi_{_{^{ik}}} = \delta_{_{^{iI}}}
\end{equation}

for $\xi$ basis the right boundary value can be very high if the last internal knot is sufficiently close to $1$
\begin{equation}
  \Pi_{_{^{1}}}^{\ast} \xi_{_{^{ik}}} = \frac{k-t_{_{^{I}}}}{1-t_{_{^{I}}}}
\end{equation}

we define new functions $\sigma_{_{^{ik}}}:= N_{_{^{ik}}} - \xi_{_{^{ik}}}$ 
in order to relate $\mathcal{G}^{\ast}$ with the former calculations, it's the same as
\begin{equation}
  \sigma_{_{^{ik}}} = - \!\ \partial^{\ast} N_{_{^{ik}}}
\end{equation}

the main set of relations is
\begin{equation}
\begin{array}{cccl}
   \left[ 1 - \mathcal{N}^{\ast} \right] & \!\!\!\! \xi_{_{^{ik}}}     & \!\!\!=\!\!\! &  \xi_{_{^{ik}}} - \delta_{_{^{iI}}} \\
   \left[ \mathcal{G}^{\ast} - 1 \right] & \!\!\!\! \sigma_{_{^{ik}}}  & \!\!\!=\!\!\! &  \xi_{_{^{ik}}} - \delta_{_{^{iI}}} \\
   \left[ \mathcal{S}^{\ast} - 1 \right] & \!\!\!\! \xi_{_{^{ik}}}     & \!\!\!=\!\!\! &  \sigma_{_{^{ik}}}
\end{array}
\label{main_set}
\end{equation}

the partition of unity property of $B$-splines (\ref{partition_unity_knots})
leads to a $\sigma$ function constraint
\begin{equation}
  \sigma_{_{^{-k+1,k}}} = -\sum_{_{^{i=-k+2}}}^{_{^{I}}} \sigma_{_{^{ik}}}
\end{equation}

with this and ({\ref{normalizacion_chi}}) formula the main set of relations is finally rewritten
\begin{equation}
\begin{array}{clclr}
   [1 - \mathcal{N}^{\ast}]
        & \!\!\!\! \xi_{_{^{\mu k}}}
        & \!\!\!\!\!=\!\!\!
        & \xi_{_{^{\mu k}}} - \delta_{_{^{\mu I}}} \sum_{_{^{\nu=-k+1}}}^{_{^{I}}} \xi_{_{^{\nu k}}}
        & \ \ \ \ \ \ \ \ \ \ \ \ \ \ \ \ \ {\textrm{\scriptsize{$\mu=-k+1,...,I$}}} \\[0.2cm]
   [\mathcal{G}^{\ast} - 1]
        & \!\!\!\! \sigma_{_{^{ik}}}
        & \!\!\!\!\!=\!\!\!
        & \xi_{_{^{ik}}} - \delta_{_{^{iI}}} \sum_{_{^{\nu=-k+1}}}^{_{^{I}}} \xi_{_{^{\nu k}}}
        & \ \ \ \ \ \ \ \ \ \ \ \ \ \ \ \ \ {\textrm{\scriptsize{$i=-k+2,...,I$}}} \\[0.2cm]
   [\mathcal{S}^{\ast} - 1]
        & \!\!\!\! \xi_{_{^{ik}}}
        & \!\!\!\!\!=\!\!\!
        & \sigma_{_{^{ik}}}&\\[0.2cm]
   [\mathcal{S}^{\ast} - 1]
        & \!\!\!\! \xi_{_{^{-k+1,k}}}
        & \!\!\!\!\!=\!\!\!
        & -\sum_{_{^{j=-k+2}}}^{_{^{I}}} \sigma_{_{^{jk}}} &
\end{array}
\end{equation}

the vector spaces spaned by these functions are
\begin{equation}
\begin{array}{ccccccl}
   H_{_{^{k}}}
      & \!\!\!\!:=\!\!\!
      & \langle\{ \xi_{_{^{ik}}}    \}_{_{^{i=-k+1}}}^{_{^{I}}} \rangle
      & \ \ \ \ \ \ \ \ \ \ \ \ \ \
      & {\textrm{dim}} \!\ H_{_{^{k}}}
      & \!\!\!\!=\!\!\!
      & I+k \\
   K_{_{^{k}}}
      & \!\!\!\!:=\!\!\!
      & \langle\{ \sigma_{_{^{ik}}} \}_{_{^{i=-k+2}}}^{_{^{I}}} \rangle
      & \ \ \ \ \ \ \ \ \ \ \ \ \ \
      & {\textrm{dim}} \!\ K_{_{^{k}}}
      & \!\!\!\!=\!\!\!
      & I+k-1
\end{array}
\end{equation}

given a number of levels $L+1=I+k$
the projection operators induce the C1 constraint (\ref{factorizationC1}) in grid-point space  
\begin{equation}
 \xymatrix @!0 @C=1.0cm @R=0.85cm{
        H_{_{^{k}}} \ar[rrrrr]^{1-\mathcal{N}^{\ast}}
              \ar[rrrd]_{\mathcal{S}^{\ast}-1}
        &&&&& H_{_{^{k}}} \ar[dddd]^{\mathcal{P}_{_{H k}}} \\
        &&& K_{_{^{k}}} \ar[rru]_{\mathcal{G}^{\ast}-1}
                \ar@<-.5ex>[dddd]_{\mathcal{P}_{_{K k}}} \\ \\ \\
        \mathcal{F}_{(0:L)} \ar@{.>}[rrrrr]^{\!\!\!\!\!\!\!\!\!\!\!\!\widehat{1-\mathcal{N}^{\ast}}}
                     \ar@{.>}[rrrd]_{\widehat{\mathcal{S}^{\ast}-1}}
                     \ar[uuuu]^{\mathcal{Q}_{_{H k}}}
        &&&&& \mathcal{F}_{(0:L)} \\
        &&& \mathcal{F}_{(1:L)} \ar@<-.5ex>[uuuu]_{\mathcal{Q}_{_{K k}}}
                        \ar@{.>}[rru]_{\widehat{\mathcal{G}^{\ast}-1}}  }
\label{c1_diagram_gp}
\end{equation}

the grid-point operators are then defined as
\begin{equation}
\begin{array}{lclcl}
   \widehat{\mathcal{N}^{\ast}}
      & \!\!\!:=\!\!\!
      & 1 - \mathcal{P}_{_{H k}} 
      & \!\!\!\!\! [1 - \mathcal{N}^{\ast}] \!\!\!\!\!
      & \mathcal{Q}_{_{H k}} \\
   \widehat{\mathcal{G}^{\ast}}
      & \!\!\!:=\!\!\!
      & 1 + \mathcal{P}_{_{H k}} 
      & \!\!\!\!\! [\mathcal{G}^{\ast} - 1] \!\!\!\!\!
      & \mathcal{Q}_{_{K k}}  \\
   \widehat{\mathcal{S}^{\ast}}
      & \!\!\!:=\!\!\!
      & 1 + \mathcal{P}_{_{K k}} 
      & \!\!\!\!\! [\mathcal{S}^{\ast} - 1] \!\!\!\!\!
      & \mathcal{Q}_{_{H k}}
\end{array}
\end{equation}

we develop their expression according to the Einstein summation convention
which states that there is an implicit summation over repeated dummy index
(e.g. $j$ in following formulae).
In a projection such that $\mathcal{Q}_{_{H}}= \mathcal{P}_{_{H}}^{_{^{-1}}}$
\begin{equation}
\begin{array}{ccccccl}
   {[}\widehat{\mathcal{N}^{\ast}}{]}_{_{^{MN}}}
      & \!\!\!\!\!=\!\!\!\!
      & 
      & \!\!\!\!\!
      & 
      & \!\!\!\!\!
      & [\mathcal{Q}_{_{H}}]_{_{^{IN}}} \\ 
   {[}\widehat{\mathcal{G}^{\ast}}{]}_{_{^{Mn}}}
      & \!\!\!\!\!=\!\!\!\!
      & \delta_{_{^{Mn}}}
      & \!\!\!\!\!+\!\!\!\!
      & [\mathcal{P}_{_{H}}]_{_{^{M j}}} [\mathcal{Q}_{_{K}}]_{_{^{j n}}}
      & \!\!\!\!\!-\!\!\!\!
      & [\mathcal{Q}_{_{K}}]_{_{^{In}}} \\ 
   {[}\widehat{\mathcal{S}^{\ast}}{]}_{_{^{mN}}}
      & \!\!\!\!\!=\!\!\!\!
      & \delta_{_{^{mN}}}
      & \!\!\!\!\!+\!\!\!\!
      & [\mathcal{P}_{_{K}}]_{_{^{mj}}} [\mathcal{Q}_{_{H}}]_{_{^{j N}}}
      & \!\!\!\!\!-\!\!\!\!
      & [\mathcal{Q}_{_{H}}]_{_{^{-k+1,N}}} \sum_{_{^{j}}} [\mathcal{P}_{_{K}}]_{_{^{mj}}}
\end{array}
\end{equation}

the indexes are
\begin{equation}
\begin{array}{ccl}
  {\textrm{\scriptsize{$M,N$}}} &
  \!\!\!{\textrm{\scriptsize{$\in$}}}\!\!\! &
  {\textrm{\scriptsize{$\{0,...,L\}$}}}\\
  {\textrm{\scriptsize{$m,n$}}} &
  \!\!\!{\textrm{\scriptsize{$\in$}}}\!\!\! &
  {\textrm{\scriptsize{$\{1,...,L\}$}}}
\end{array}
\ \ \ \ \ \ \ \ \ \ \ \ \ \
\begin{array}{ccl}
  {\textrm{\scriptsize{$\mu$}}} &
  \!\!\!{\textrm{\scriptsize{$\in$}}}\!\!\! &
  {\textrm{\scriptsize{$\{-k\!+\!1,...,I\}$}}}\\
  {\textrm{\scriptsize{$j$}}} &
  \!\!\!{\textrm{\scriptsize{$\in$}}}\!\!\! &
  {\textrm{\scriptsize{$\{-k\!+\!2,...,I\}$}}}
\end{array}
\end{equation}

setting the level $t(0)=0$ we consider the boundary conditions
\begin{equation}
\begin{array}{lcl}
  {[}\mathcal{P}_{_{H}}{]}_{_{^{0\mu}}}
     & \!\!\!\!\!=\!\!\!\!
     & \Pi_{_{^{0}}}^{\ast} \xi_{_{^{\mu k}}} = t\partial N_{_{^{\mu k}}} |_{_{t=0}} + \Pi_{_{^{0}}}^{\ast} N_{_{^{\mu k}}} = \delta_{_{^{\mu,-k+1}}} \\
  {[}\mathcal{Q}_{_{H}}{]}_{_{^{-k+1,N}}}
     & \!\!\!\!\!=\!\!\!\!
     & \delta_{_{^{N0}}}
\end{array}
\end{equation}

to work with a set of equations restricted to $m,n$ indices
we extrapolate functions in the boundary $f_{_{^{0}}}=\alpha_{_{^{n}}} f_{_{^{n}}}$ being $\sum_{_{^{n}}} \alpha_{_{^{n}}}=1$
\begin{equation}
\boxed{
\begin{array}{lcl}
   {[}\widehat{\mathcal{N}^{\ast}}{]}_{_{^{n}}}
      & \!\!\!\!\!=\!\!\!\!
      & [\mathcal{Q}_{_{H}}]_{_{^{In}}} + \alpha_{_{^{n}}} [\mathcal{Q}_{_{H}}]_{_{^{I0}}} \\
   {[}\widehat{\mathcal{G}^{\ast}}{]}_{_{^{mn}}}
      & \!\!\!\!\!=\!\!\!\!
      & \delta_{_{^{mn}}} + [\mathcal{Q}_{_{K}}]_{_{^{jn}}} \left\{ [\mathcal{P}_{_{H}}]_{_{^{mj}}} - \delta_{_{^{I j}}} \right\} \\
   {[}\widehat{\mathcal{S}^{\ast}}{]}_{_{^{mn}}}
      & \!\!\!\!\!=\!\!\!\!
      & \delta_{_{^{mn}}} + [\mathcal{P}_{_{K}}]_{_{^{mj}}} 
        \left\{ [\mathcal{Q}_{_{H}}]_{_{^{jn}}} + \alpha_{_{^{n}}} \left[ [\mathcal{Q}_{_{H}}]_{_{^{j0}}} - 1 \right] \right\}
\end{array}}
\end{equation}

the condition $\widehat{\mathcal{S}^{\ast}} 1=1$ and $\widehat{\mathcal{N}^{\ast}} 1=1$ 
are guaranteed by the \ref{sum1} property analog with $B$-splines


\section{Conclusion}
The finite element version of the vertical operators of the nh-model can be constructed
starting from the prerequisite of satisfy some kinds of constraints as shown in the present paper. 
However, there is still the problem of the choice of internal knots, 
this is of primary importance in order to reduce the noise produced in the model by the discretized operators. 
As pointed out by J. Vivoda they should be such that the maxima of the associated splines be as close of full levels as possible
in order to work with diagonal dominant projection operators which are suitable to invert. 
The coding of these operators in the model should be studied carefully in order to keep the stability and the skill of the model, 
this will be developed in future work.

\section*{Acknowledgements}
The author would like to be grateful for to work to Mariano Hortal (AEMET),
for his continuous support and for having the opportunity of learn with him
about the dynamics of the ALADIN-HIRLAM nwp system, his knowledge of the model is inmense.
The author also wants to acknowledge the discussions during stays at Prague and Valencia
with Petra Smolíková (CHMI), Jozef Vivoda (SHMÚ), and Juan Simarro (AEMET)
about different approaches on setting a finite element version of the vertical operators of the model

\section{Appendix}
\label{cap:appendix}

\subsection{Proofs on B-splines integral formulae}
\label{cap:proofs_integrals}
In this section we develop the integral formula for $B$-splines 
in the case of the specific boundaries of integration that define $\mathcal{I}$, $\mathcal{J}$, $\mathcal{N}$ operators.
The general formula can be found in \citealp{dBLS}

\begin{proposition}
the operator $\mathcal{I}$ acts on $B$-splines as
\begin{equation}
  \mathcal{I} N_{_{^{ik}}} = -\frac{\Del{ik}}{k} \sum_{_{^{s=-\infty}}}^{_{^{i-1}}} N_{_{^{s,k+1}}} 
\label{int_i}
\end{equation} 
\end{proposition}

the integral of the generatrix functions is
\begin{equation}
  \int^{_{^{x=s}}}_{_{^{x=t}}} dx \!\ g_{_{^{k}}} (s;x) = \frac{1}{k} g_{_{^{k+1}}} (s;t) 
\end{equation}

this has the same value if ve choose the interval of integration $[t,s']$ instead of $[t,s]$ for $s'>s$.
In particular if we set $s'=t_{\infty}$ results $\mathcal{I} g_{_{^{k}}} (s;\cdot) = -\frac{1}{k} g_{_{^{k+1}}} (s;\cdot)$.
Taking divided differences on $s$ and then the difference between $\mathcal{I} M_{_{^{ik}}}$ and $\mathcal{I} M_{_{^{i-1,k}}}$
\begin{equation}
  \mathcal{I} M_{_{^{ik}}} = \mathcal{I} M_{_{^{i-1,k}}} - \frac{1}{k} N_{_{^{i-1,k+1}}}
\end{equation}
 
for $j<i$
\begin{equation}
  \mathcal{I} M_{_{^{ik}}} = \mathcal{I} M_{_{^{jk}}} - \frac{1}{k} \sum_{_{^{r=j}}}^{_{^{i-1}}} N_{_{^{r,k+1}}}
\end{equation}

to end the proof we need to prove that $\forall t \in (-t_{_{^{\infty}}},t_{_{^{\infty}}})$
\begin{equation}
  \lim_{_{^{j\rightarrow -\infty}}} \int^{_{^{x=t}}}_{_{^{x=t_{\infty}}}} dx \!\ M_{_{^{jk}}}(x) = 0
\end{equation}

if $t_{_{^{-\infty}}}>-\infty$ let $t\in (-t_{_{^{\infty}}},t_{_{^{\infty}}})$, 
by the definition of limit given $\varepsilon>0$ there exists $N$ such that $| t_{_{^{-\infty}}} \!\! - t_{_{^{n}}} | < \varepsilon$
for all $n\leq N$, we choose any value $0<\varepsilon < | t_{_{^{-\infty}}} \!\! - t \!\ |$
and knots $\{ t_{_{^{N_{\varepsilon}-k}}},..., t_{_{^{N_{\varepsilon}}}} \}$  so that $t$
does not belong to the support of $M_{_{^{N_{\varepsilon}-k,k}}}$
and then to the support of $M_{_{^{nk}}}$\\

in the case with $t_{_{^{-\infty}}}=-\infty$,
given $t$ it exists $N$ such that $t_{_{^{n}}}<t$ for all $n\leq N$, the rest of the proof is like the former case
\qed\\

\begin{theorem}
the integral of $B$-splines over the whole domain is
\begin{equation}
  \mathcal{N} N_{_{^{ik}}} = \frac{\Del{ik}}{k} 
\end{equation}
\end{theorem}

we apply (\ref{int_i}) in the limit $t\rightarrow t_{_{^{-\infty}}}$
\begin{equation}
  \int_{_{^{t_{-\infty}}}}^{_{^{t_{\infty}}}} N_{_{^{ik}}} \!\ dt = \frac{\Del{ik}}{k}
  \lim_{_{^{t\rightarrow t_{-\infty}}}} \sum_{_{^{s=-\infty}}}^{_{^{i-1}}} N_{_{^{s,k+1}}}(t)
\end{equation}

by the partition of unity property we can reduce our proof to the following formula for all $i\in\mathbb{Z}$ 
\begin{equation}
  \lim_{_{^{t\rightarrow t_{-\infty}}}} \Del{ik}  \sum_{_{^{s=i}}}^{_{^{\infty}}} N_{_{^{s,k+1}}}(t) = 0
\end{equation}

if $t_{_{^{-\infty}}}\notin\mathtt{t}$ it is sufficient to take $t<t_{i}$. 
In the other case with $t_{_{^{-\infty}}}\in\mathtt{t}$ there exists $N$ such that $t_{_{^{n}}}=t_{_{^{-\infty}}}$
for all $n\leq N$ and $t_{_{^{-\infty}}}<t_{_{^{N+1}}}$, so we can write
\begin{equation}
  \lim_{_{^{t\rightarrow t_{-\infty}}}} \Del{ik} \sum_{_{^{s=i}}}^{_{^{\infty}}} N_{_{^{s,k+1}}}(t) 
                                      = \Del{ik} \sum_{_{^{s=i}}}^{_{^{\infty}}} N_{_{^{s,k+1}}}(t_{_{^{N}}})
\end{equation}

for $N<i$ this has zero value, for $i\leq N-k$ the step $\Del{ik}$ is also zero. 
Finally for $N-K+1\leq i\leq N$ we apply (\ref{recursive_b_splines})  
to get the auxiliar formula
\begin{equation}
  N_{_{^{il}}} (t_{_{^{N}}}) = \frac{\Del{il}}{\Del{i+1,l-1}} N_{_{^{i+1,l-1}}} (t_{_{^{N}}})
\end{equation}
reindexing $i=N-r$ where $r=0,...,k-1$ we write
$N_{_{^{N-r,k+1}}} (t_{_{^{N}}})= {\textrm{\scriptsize{cte. }}} N_{_{^{N+1,k-r}}} (t_{_{^{N}}})=0$
being $k-r \geq 1$
\qed\\

\subsection{Some analytical properties on vertical operators}
\label{cap:vertical_operators}
\label{cap:vertical_operators2}

In this appendix we present some properties of the 
vertical operators defined in the previous section \ref{cap:vertical_operators1}.
The matrices involved in the constraints present in the linear systems are \citep{Bb}
\begin{equation}
\begin{array}{lcl}
  \mathcal{A}^{\ast}_{_{^{1}}} 
    & \!\!\!:=\!\!\! 
    &  \mathcal{G}^{\ast} \mathcal{S}^{\ast} - \mathcal{G}^{\ast} - \mathcal{S}^{\ast} + \mathcal{N}^{\ast} \\[0.2cm]
  \mathcal{A}^{\ast}_{_{^{2}}} 
    & \!\!\!:=\!\!\! 
    &  \mathcal{S}^{\ast} \mathcal{G}^{\ast} - \frac{c_{pd}}{c_{vd}} [ \mathcal{G}^{\ast} + \mathcal{S}^{\ast} ]
\end{array}
\end{equation}

that satisfy the constraints C1 and C2
\begin{equation}
   \mathcal{A}^{\ast}_{_{^{1}}} = 0
   \ \ \ \ \ \ \ \ \ \ \ \ \ \ \ \ \ \ \ \ \ \ \ \ \ \ \ \
   \mathcal{L}^{\ast} \mathcal{A}^{\ast}_{_{^{2}}} = \frac{c_{pd}}{c_{vd}} - 1 
\end{equation}


the boundary conditions of the vertical operators 
\begin{equation}
 \xymatrix @C=1.2cm @R=0.2cm{ 
   \mathcal{G}^{\ast} f  \ar[r]^{t \rightarrow 0} & \mathcal{N}^{\ast} \frac{f}{t} \\
   \mathcal{G}^{\ast} f  \ar[r]^{t \rightarrow 1} & 0 \ \ \ \ }
 \ \ \ \ \ \ \ \ \ \ \ \ \ \ \
 \xymatrix @C=1.2cm @R=0.2cm{
   \mathcal{S}^{\ast} f  \ar[r]^{t \rightarrow 0} & f(0) \\
   \mathcal{S}^{\ast} f  \ar[r]^{t \rightarrow 1} & \mathcal{N}^{\ast} f }
\end{equation}

acting on constants $\mathcal{S}^{\ast} \lambda=\lambda$ and $\mathcal{N}^{\ast} \lambda=\lambda$.
Some useful relations between operators are
\begin{equation}
  \begin{array}{lcl}
     \partial^{\ast} \!\ \mathcal{S}^{\ast}  & \!\!\!\!\!=\!\!\!\! & 1 -\mathcal{S}^{\ast} \\
     \partial^{\ast} \!\ \mathcal{G}^{\ast}  & \!\!\!\!\!=\!\!\!\! & -1 \\
     \partial^{\ast} \!\ \mathcal{N}^{\ast}  & \!\!\!\!\!=\!\!\!\! & 0
  \end{array}
  \ \ \ \ \ \ \ \ 
  \begin{array}{lcl}
     \mathcal{L}^{\ast} \!\ \mathcal{S}^{\ast}  & \!\!\!\!\!=\!\!\!\! & \partial^{\ast} \\
     \mathcal{L}^{\ast} \!\ \mathcal{G}^{\ast}  & \!\!\!\!\!=\!\!\!\! & -(\partial^{\ast} \!+1) \\
     \mathcal{L}^{\ast} \!\ \mathcal{N}^{\ast}  & \!\!\!\!\!=\!\!\!\! & 0
  \end{array}
  \ \ \ \ \ \ \ \ 
  \begin{array}{lcl}
    \mathcal{S}^{\ast} \!\ \partial^{\ast}     & \!\!\!\!\!=\!\!\!\! &  1 - \mathcal{S}^{\ast} \\
    \mathcal{S}^{\ast} \!\ \mathcal{L}^{\ast}  & \!\!\!\!\!=\!\!\!\! &  \partial^{\ast} \\
    \mathcal{S}^{\ast} \!\ \mathcal{G}^{\ast}  & \!\!\!\!\!=\!\!\!\! &  \mathcal{S}^{\ast} \!+ \mathcal{G}^{\ast}
  \end{array}
  \ \ \ \ \ \ \ \ 
  \begin{array}{lcl}
    \left[ \mathcal{S}^{\ast},     \partial^{\ast} \right]  & \!\!\!\!\!=\!\!\!\! &  0 \\
    \left[ \mathcal{S}^{\ast},  \mathcal{G}^{\ast} \right]  & \!\!\!\!\!=\!\!\!\! &  \mathcal{N}^{\ast} \\
    \mathcal{L}^{\ast} \!\ \mathcal{S}^{\ast} \!\ \mathcal{G}^{\ast} & \!\!\!\!\!=\!\!\!\! & -1 
  \end{array}
\end{equation}

defining the operator $\Pi_{x}^{\ast}$ which evaluates the value of a given function in a point $x$ we have
\begin{equation}
  \begin{array}{ccl}
    \mathcal{G}^{\ast} \!\ \partial^{\ast} & \!\!\!\!\!=\!\!\!\! & \Pi_{_{^{1}}}^{\ast} - 1 \\ 
    \mathcal{N}^{\ast} \!\ \partial^{\ast} & \!\!\!\!\!=\!\!\!\! & \Pi_{_{^{1}}}^{\ast} - \mathcal{N}^{\ast}
  \end{array}
  \ \ \ \ \ \ \ \
  \begin{array}{ccl}
    &\!\!\!\!\!&\\
    \mathcal{N}^{\ast} \!\ (\partial^{\ast}\!+1) & \!\!\!\!\!=\!\!\!\! & \Pi_{_{^{1}}}^{\ast}
  \end{array}
  \ \ \ \ \ \ \ \
  \begin{array}{ccl}
    \mathcal{G}^{\ast} \!\ \mathcal{L}^{\ast}  & \!\!\!\!\!=\!\!\!\! & (\Pi_{_{^{1}}}^{\ast} - 1) (\partial^{\ast}\!+1) \\
    \mathcal{N}^{\ast} \!\ \mathcal{L}^{\ast}  & \!\!\!\!\!=\!\!\!\! & \Pi_{_{^{1}}}^{\ast} \partial^{\ast}
  \end{array}
  \ \ \ \ \ \ \ \
  \begin{array}{ccl}
    \Pi_{_{^{1}}}^{\ast} \mathcal{G}^{\ast} & \!\!\!\!\!=\!\!\!\! & 0 \\
    \Pi_{_{^{1}}}^{\ast} \mathcal{S}^{\ast} & \!\!\!\!\!=\!\!\!\! & \mathcal{N}^{\ast} 
  \end{array}
\label{set_pi1}
\end{equation}

an important fact is that $\partial^{\ast} \!+ 1$ is the inverse operator of $\mathcal{S}^{\ast}$
\begin{equation}
  \mathcal{S}^{\ast\!\ -1} = \partial^{\ast} \!+ 1
\label{SDinverse}
\end{equation}

if we consider the $L^{2}$ scalar product
$ \langle f | g \rangle := \int^{_{^{1}}}_{^{_{0}}} f g \!\ dt$
the operators $\mathcal{G}^{\ast}$ and $\mathcal{S}^{\ast}$ are adjoint. Some self-adjoint operators are
$\mathcal{N}^{\ast}, \mathcal{G}^{\ast} \! +\mathcal{S}^{\ast}, \mathcal{G}^{\ast}\mathcal{S}^{\ast},
 \mathcal{S}^{\ast}\mathcal{G}^{\ast}, \mathcal{A}^{\ast}_{_{^{2}}} $

\subsection{Basis functions}
\label{cap:basis}








In this appendix we show the explicit formulae that are used for computations.
The basis that are used to solve C1 constraint and
$\omega_{_{^{ik}}} := -\mathcal{L}^{\ast} \!\ N_{_{^{ik}}}$
have the explicit expression
\begin{equation}
   \xi_{_{^{ik}}} =  
       ( kt -t_{_{^{i}}} )     \!\ \frac{N_{_{^{i,k-1}}}}{\Delta_{_{^{i,k-1}}}} 
     + ( t_{_{^{i+k}}}\!\!-kt) \!\ \frac{N_{_{^{i+1,k-1}}}}{\Delta_{_{^{i+1,k-1}}}}  
\end{equation}

\begin{equation}
  \sigma_{_{^{ik}}} = t(k-1) \left[
       - \frac{N_{_{^{i,k-1}}}}  {\Delta_{_{^{i,k-1}}}}
       + \frac{N_{_{^{i+1,k-1}}}}{\Delta_{_{^{i+1,k-1}}}} \right]
\end{equation}

\begin{equation}
 \omega_{_{^{ik}}} = t(k-1) \left[
         \frac{2t_{_{^{i}}}-kt}{\Delta_{_{^{i,k-1}}}} \ \frac{N_{_{^{i,k-2}}}}  {\Delta_{_{^{i,k-2}}}}
       + \left( \frac{kt-2t_{_{^{i+k-1}}}}{\Delta_{_{^{i,k-1}}}} + \frac{kt-2t_{_{^{i+1}}}}{\Delta_{_{^{i+1,k-1}}}} \right)
         \frac{N_{_{^{i+1,k-2}}}}{\Delta_{_{^{i+1,k-2}}}}
       + \frac{2t_{_{^{i+k}}}-kt}{\Delta_{_{^{i+1,k-1}}}} \  \frac{N_{_{^{i+2,k-2}}}}  {\Delta_{_{^{i+2,k-2}}}}
         \right]
\end{equation}

the support of the basis functions is related to knots, for them is enough to apply their respective definitions
\begin{equation}
  {\textrm{supp}} \!\      N_{_{^{ik}}} =
  {\textrm{supp}} \!\    \xi_{_{^{ik}}} =
  {\textrm{supp}} \!\ \sigma_{_{^{ik}}} =
  {\textrm{supp}} \!\ \omega_{_{^{ik}}} =
  [ t_{_{^{i}}}, t_{_{^{i+k}}} ]
\end{equation}

we write relations between the basis diagrammatically
\begin{equation}
 \xymatrix @C=2.4cm @R=2.0cm{
   \boxed{\sigma_{_{^{ik}}}}
      \ar@<.5ex>[r]^{\mathcal{G}^{\ast}} 
      \ar@<.5ex>[d]^{\partial^{\ast}\!+1} &
   \boxed{N_{_{^{ik}}} \!\!- \delta_{_{^{iI}}}} 
      \ar@<.5ex>[l]^{-\partial^{\ast}}
      \ar@<.5ex>[d]^{\partial^{\ast}\!+1} 
      \ar@<.5ex>[dl]^{-\mathcal{L}^{\ast}}  \\ 
   \boxed{\omega_{_{^{ik}}}}
      \ar@<.5ex>[r]^{\mathcal{G}^{\ast}\!-\mathcal{N}^{\ast}}
      \ar@<.5ex>[u]^{\mathcal{S}^{\ast}} 
      \ar@<.5ex>[ur]^{\mathcal{G}^{\ast}\mathcal{S}^{\ast}} &
   \boxed{\xi_{_{^{ik}}} \!\!- \delta_{_{^{ik}}}}  
      \ar@<.5ex>[l]^{-\partial^{\ast}}
      \ar@<.5ex>[u]^{\mathcal{S}^{\ast}}  }
\end{equation}

to compute projection matrix using the Sobolev inner product \citep{O'S}
we present the derivatives of the basis functions,
the relation $\partial \!\ \partial^{\ast}= (\partial^{\ast}\!+1) \!\ \partial$
and its generalization $[ \partial^{_{^{n}}}, \partial^{\ast} ] = n\!\ \partial^{_{^{n}}}$ 
allows to compute first derivatives
\begin{equation}
\begin{array}{cccccc}
  \partial \!\ N_{_{^{ik}}} 
    & \!\!\!=\!\!\! 
    & (k-1) [ \!\!\!\!\! 
    & \tilde{N}_{_{^{i,k-1}}} 
    & \!\!\!\!\!\!-\!\!\!\!\!
    & \tilde{N}_{_{^{i+1,k-1}}} ] \\
  \partial \!\ \xi_{_{^{ik}}}
    & \!\!\!=\!\!\!
    & (k-1) [ \!\!\!\!\!
    & \tilde{\xi}_{_{^{i,k-1}}}
    & \!\!\!\!\!\!-\!\!\!\!\!
    & \tilde{\xi}_{_{^{i+1,k-1}}} ] \\
  \partial \!\ \sigma_{_{^{ik}}}
    & \!\!\!=\!\!\!
    & (k-1) [ \!\!\!\!\!
    & \tilde{\sigma}_{_{^{i,k-1}}}
    & \!\!\!\!\!\!-\!\!\!\!\!
    & \tilde{\sigma}_{_{^{i+1,k-1}}} ] \\
  \partial \!\ \omega_{_{^{ik}}}
    & \!\!\!=\!\!\!
    & (k-1) [ \!\!\!\!\!
    & \tilde{\omega}_{_{^{i,k-1}}}
    & \!\!\!\!\!\!-\!\!\!\!\!
    & \tilde{\omega}_{_{^{i+1,k-1}}} ] 
\end{array}
\end{equation}

where we had defined the auxiliar basis
\begin{equation}
\begin{array}{lcc}
  \tilde{N}_{_{^{ik}}}      &\!\!\!\!:=\!\!\!& \frac{N_{_{^{ik}}}}{\Delta_{_{^{ik}}}} \\[0.2cm]
  \tilde{\xi}_{_{^{ik}}}    &\!\!\!\!:=\!\!\!& \frac{\xi_{_{^{ik}}}    + N_{_{^{ik}}}}{\Delta_{_{^{ik}}}} \\[0.2cm]
  \tilde{\sigma}_{_{^{ik}}} &\!\!\!\!:=\!\!\!& \frac{\sigma_{_{^{ik}}} - N_{_{^{ik}}}}{\Delta_{_{^{ik}}}} \\[0.2cm]
  \tilde{\omega}_{_{^{ik}}} &\!\!\!\!:=\!\!\!& \frac{\omega_{_{^{ik}}} - 2\!\  N_{_{^{ik}}}}{\Delta_{_{^{ik}}}}
\end{array}
\end{equation}

and the second derivatives
\begin{equation}
\begin{array}{c}
 \partial^{2} \!\ N_{_{^{ik}}} = (k-1)(k-2) \left[
         \frac{\hat{N}_{_{^{i,k-2}}}}{\Delta_{_{^{i,k-1}}}}
       - \left( \frac{1}{\Delta_{_{^{i,k-1}}}} + \frac{1}{\Delta_{_{^{i+1,k-1}}}} \right)
         \hat{N}_{_{^{i+1,k-2}}}
       + \frac{\hat{N}_{_{^{i+2,k-2}}}}{\Delta_{_{^{i+1,k-1}}}} 
         \right]\\
 \partial^{2} \!\ \xi_{_{^{ik}}} = (k-1)(k-2) \left[
         \frac{\hat{\xi}_{_{^{i,k-2}}}}{\Delta_{_{^{i,k-1}}}}
       - \left( \frac{1}{\Delta_{_{^{i,k-1}}}} + \frac{1}{\Delta_{_{^{i+1,k-1}}}} \right)
         \hat{\xi}_{_{^{i+1,k-2}}}
       + \frac{\hat{\xi}_{_{^{i+2,k-2}}}}{\Delta_{_{^{i+1,k-1}}}}
         \right]\\
 \partial^{2} \!\ \sigma_{_{^{ik}}} = (k-1)(k-2) \left[
         \frac{\hat{\sigma}_{_{^{i,k-2}}}}{\Delta_{_{^{i,k-1}}}}
       - \left( \frac{1}{\Delta_{_{^{i,k-1}}}} + \frac{1}{\Delta_{_{^{i+1,k-1}}}} \right)
         \hat{\sigma}_{_{^{i+1,k-2}}}
       + \frac{\hat{\sigma}_{_{^{i+2,k-2}}}}{\Delta_{_{^{i+1,k-1}}}}
         \right]\\
 \partial^{2} \!\ \omega_{_{^{ik}}} = (k-1)(k-2) \left[
         \frac{\hat{\omega}_{_{^{i,k-2}}}}{\Delta_{_{^{i,k-1}}}}
       - \left( \frac{1}{\Delta_{_{^{i,k-1}}}} + \frac{1}{\Delta_{_{^{i+1,k-1}}}} \right)
         \hat{\omega}_{_{^{i+1,k-2}}}
       + \frac{\hat{\omega}_{_{^{i+2,k-2}}}}{\Delta_{_{^{i+1,k-1}}}}
         \right]
\end{array}
\end{equation}

where
\begin{equation}
\begin{array}{lcc}
  \hat{N}_{_{^{ik}}}      &\!\!\!\!:=\!\!\!& \frac{N_{_{^{ik}}}}{\Delta_{_{^{ik}}}} \\[0.2cm]
  \hat{\xi}_{_{^{ik}}}    &\!\!\!\!:=\!\!\!& \frac{\xi_{_{^{ik}}}    + 2\!\  N_{_{^{ik}}}}{\Delta_{_{^{ik}}}} \\[0.2cm]
  \hat{\sigma}_{_{^{ik}}} &\!\!\!\!:=\!\!\!& \frac{\sigma_{_{^{ik}}} - 2\!\  N_{_{^{ik}}}}{\Delta_{_{^{ik}}}} \\[0.2cm]
  \hat{\omega}_{_{^{ik}}} &\!\!\!\!:=\!\!\!& \frac{\omega_{_{^{ik}}} + 4\!\ \sigma_{_{^{ik}}} - 6\!\  N_{_{^{ik}}}}{\Delta_{_{^{ik}}}} 
\end{array}
\end{equation}


\bibliography{thebib}
\bibliographystyle{plainnat}

\end{document}